\documentclass{iaus}
\usepackage{graphicx}

\title[Cadence Optimisation]
{Cadence Optimisation and Exoplanetary Parameter Sensitivity}

\author[Stephen R. Kane, Eric B. Ford, \& Jian Ge]
{Stephen R. Kane, Eric B. Ford, \and Jian Ge}

\affiliation{Department of Astronomy, University of Florida, 211 Bryant Space
Science Center, Gainesville, FL 32611-2055, U.S.A.}

\pubyear{2008}
\volume{249}  
\pagerange{119--126}
\begin{document}

\maketitle

\begin{abstract}
To achieve maximum planet yield for a given radial velocity survey, the
observing strategy must be carefully considered. In particular, the adopted
cadence can greatly affect the sensitivity to exoplanetary parameters such
as period and eccentricity. Here we describe simulations which aim to
maximise detections based upon the target parameter space of the survey.
\end{abstract}

\firstsection

\section{Introduction}

Large-scale radial velocity surveys for extra-solar planets require a
great deal of planning, particularly in terms of instrument considerations
and the selection of targets. The duration of the survey will affect the
sensitivity of the survey to different regions of period space.
Additionally, the cadence of the observations affect the detection of
short and long period planets and the overall planet yield. We present
simulations of different observing strategies and demonstrate the change
in sensitivity to a planetary period, mass, and eccentricity. These
results are used to calculate the relative frequency of planet detections
for various ranges of orbital parameters. By simulating a selection of
cadence configurations, the optimal cadence for a given survey duration
and observing constraints is estimated. The techniques presented here may
be applied to a wide range of planet surveys with limited resources in
order to maximise planet yield.

\section{Simulation Framework}

To investigate the detection efficiency properties of various radial
velocity observing programs, we constructed a suite of simulated datasets
using a FORTRAN code which also performs the analysis, as described by
\cite[Kane, Schneider, \& Ge (2007)]{kane07}. The parameters of the
initial simulation are based upon those of the Multi-object APO
Radial-Velocity Exoplanet Large-Area Survey (MARVELS)
(\cite[Ge et al. 2006]{ge06}). The stellar properties were estimated from
Tycho-2 stars selected for observation by MARVELS for 60 separate fields.
The noise model for the simulation was produced from the current and
planned performance of the instrument and used to generate the radial
velocity data. The cadences were defined by the number of observations
per month during bright time. An example cadence can be conveniently
expressed as 888, meaning 8 observations per month for 3 consecutive months.
Since we are concerned with exoplanet parameter sensitivity, each dataset
was injected with a planetary signature, the parameters of which were
randomly chosen from a uniform distribution including mass, period, and
eccentricity. In this way, $>$~4 million datasets were produced for each
cadence simulation. Figure 1 represents an example simulated dataset,
showing the radial velocity amplitudes and noise model.

\begin{figure}
  \hspace{2.5cm}
  \includegraphics[angle=270,width=8.2cm]{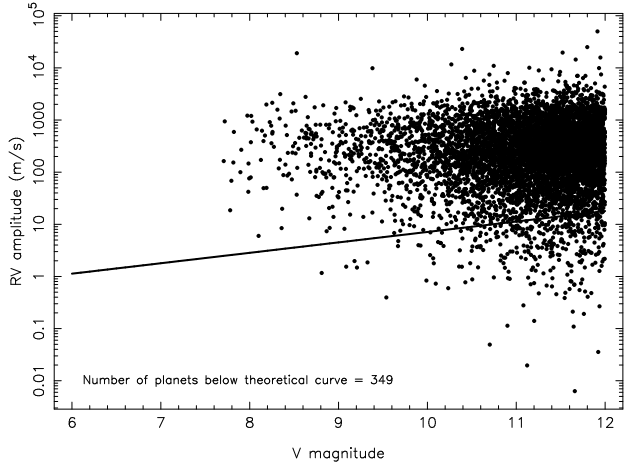}
  \caption{Radial velocity amplitudes of the planetary signatures along with
    the noise model for a small sample of datasets.}
\end{figure}

\section{Detection Criteria}

The code written for the task of sifting planet candidates from the data
uses a weighted Lomb-Scargle (L-S) periodogram to detect a periodic
signal. The number of false detections resulting from this technique
depends upon the periodic false-alarm probability threshold one adopts as
the detection criteria. We selected this threshold for each cadence by
producing a large number of datasets with no planets injected, then
executing a Monte-Carlo simulation to determine that threshold which
yields the required maximum number of false detections. In addition, we
distinguish between those detections with unique and ambiguous periods
based on the number of significant peaks in the periodogram.

\begin{figure*}
  \begin{center}
    \begin{tabular}{cc}
      \includegraphics[angle=270,width=6.2cm]{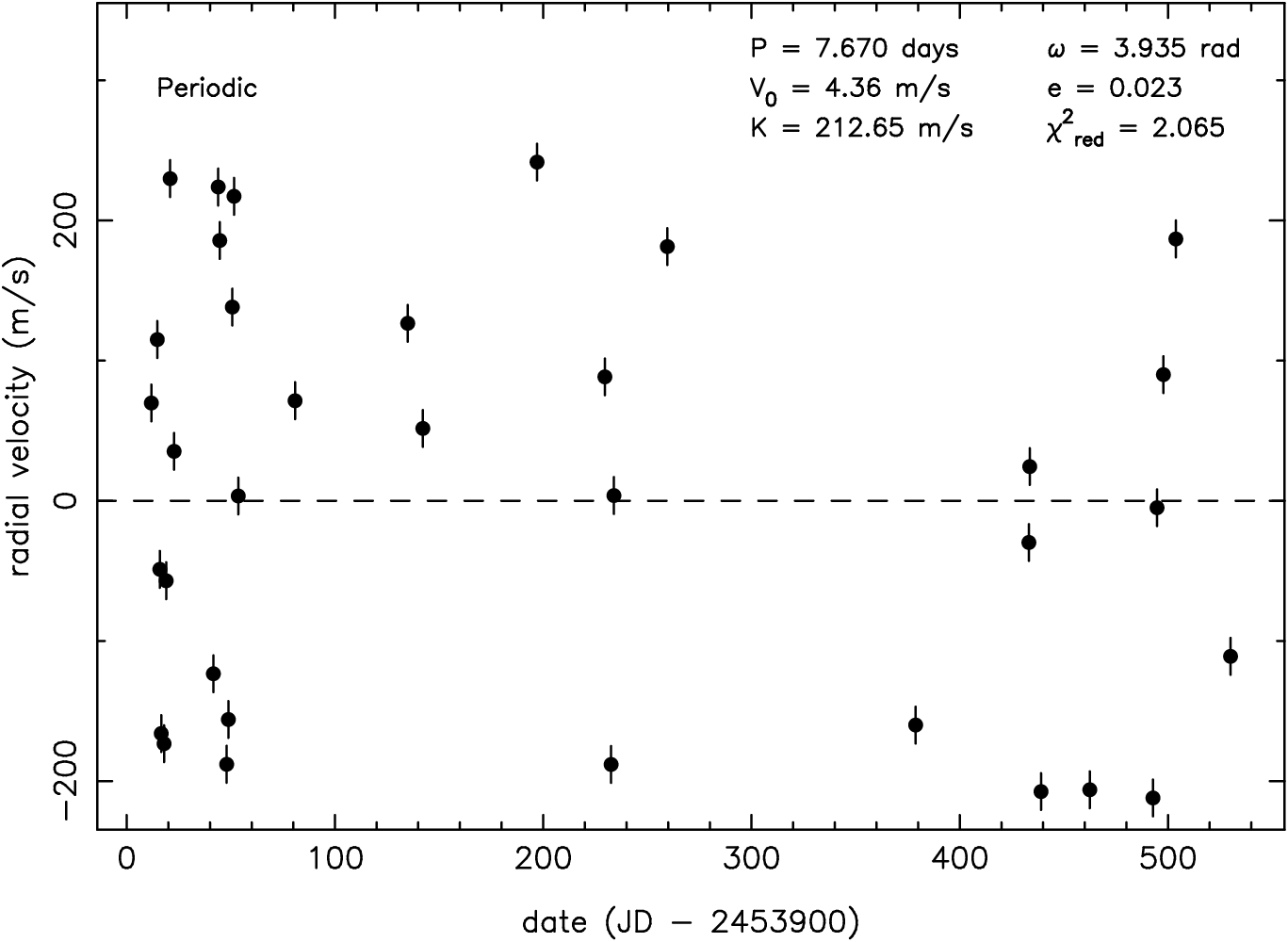} &
      \hspace{0.3cm}
      \includegraphics[angle=270,width=6.4cm]{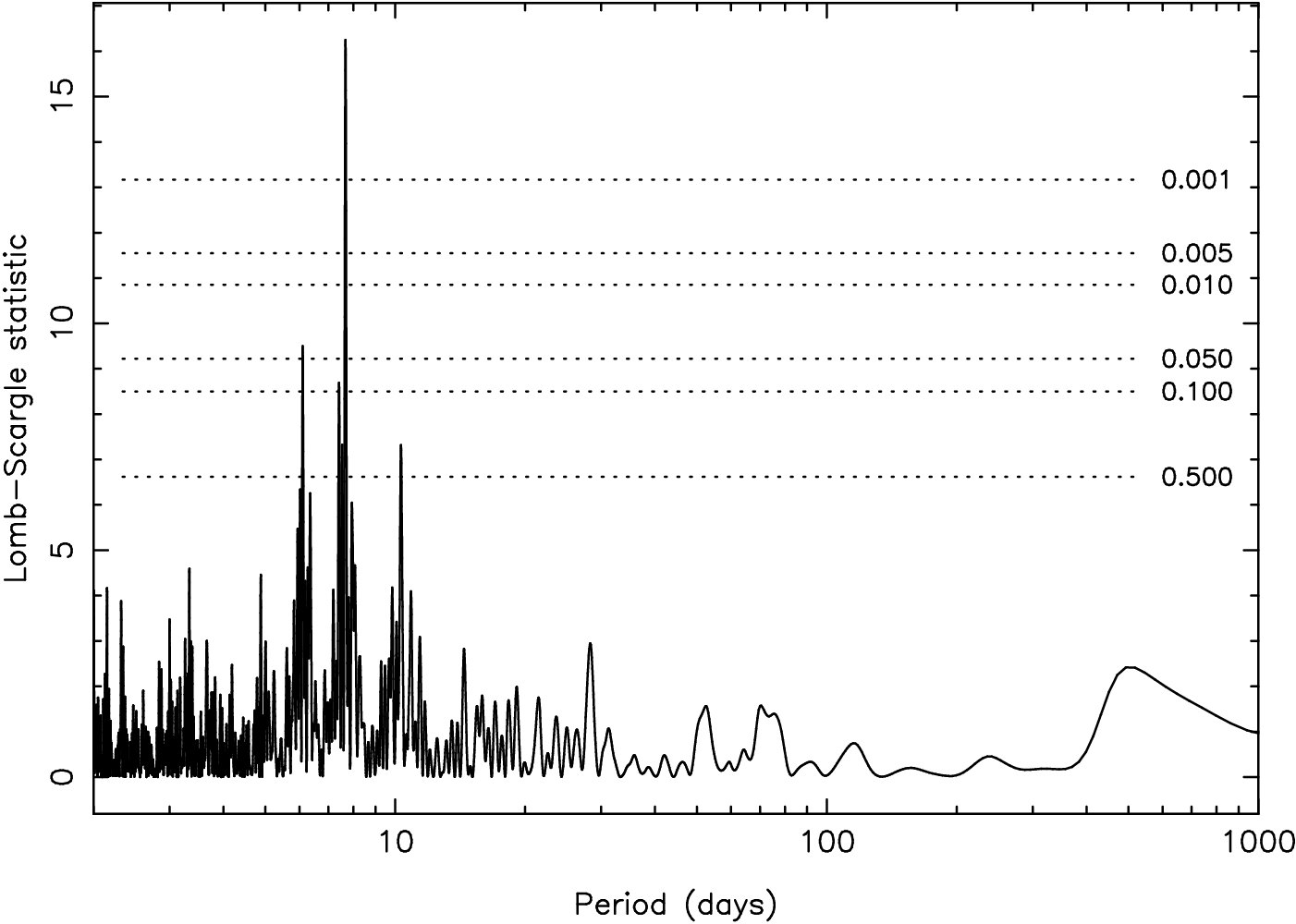} \\
    \end{tabular}
  \end{center}
  \caption{A simulated dataset for an 18 month cadence configuration (left)
    along with the accompanying periodogram (right).}
\end{figure*}

Shown in Figure 2 (left) is a typical dataset from an 18 month cadence
simulation. Figure 2 (right) also shows the corresponding periodogram
where the dotted lines indicate various false-alarm probabilities.

\section{Cadence Results and Planet Yield}

For each cadence configuration, the simulated datasets were passed through
the described detection algorithm and the results sorted by period,
eccentricity, and sensitivity (K/$\sigma$). The results for the 888 simulation
are shown in Figure 3, where the dashed line indicates all detections and the
solid line indicates only unique period detections. The four plots on the
left show the period dependence for a circular orbit and the four plots on the
right show the eccentricity dependence for a period range of 7--15 days.
The detection efficiency of the 888 cadence performs moderately well for
short-period planets with relatively circular orbits, but suffers greatly in
the long-period and mid-high eccentricity regimes.

\begin{figure*}
  \begin{center}
    \begin{tabular}{cc}
      \includegraphics[angle=270,width=6.3cm]{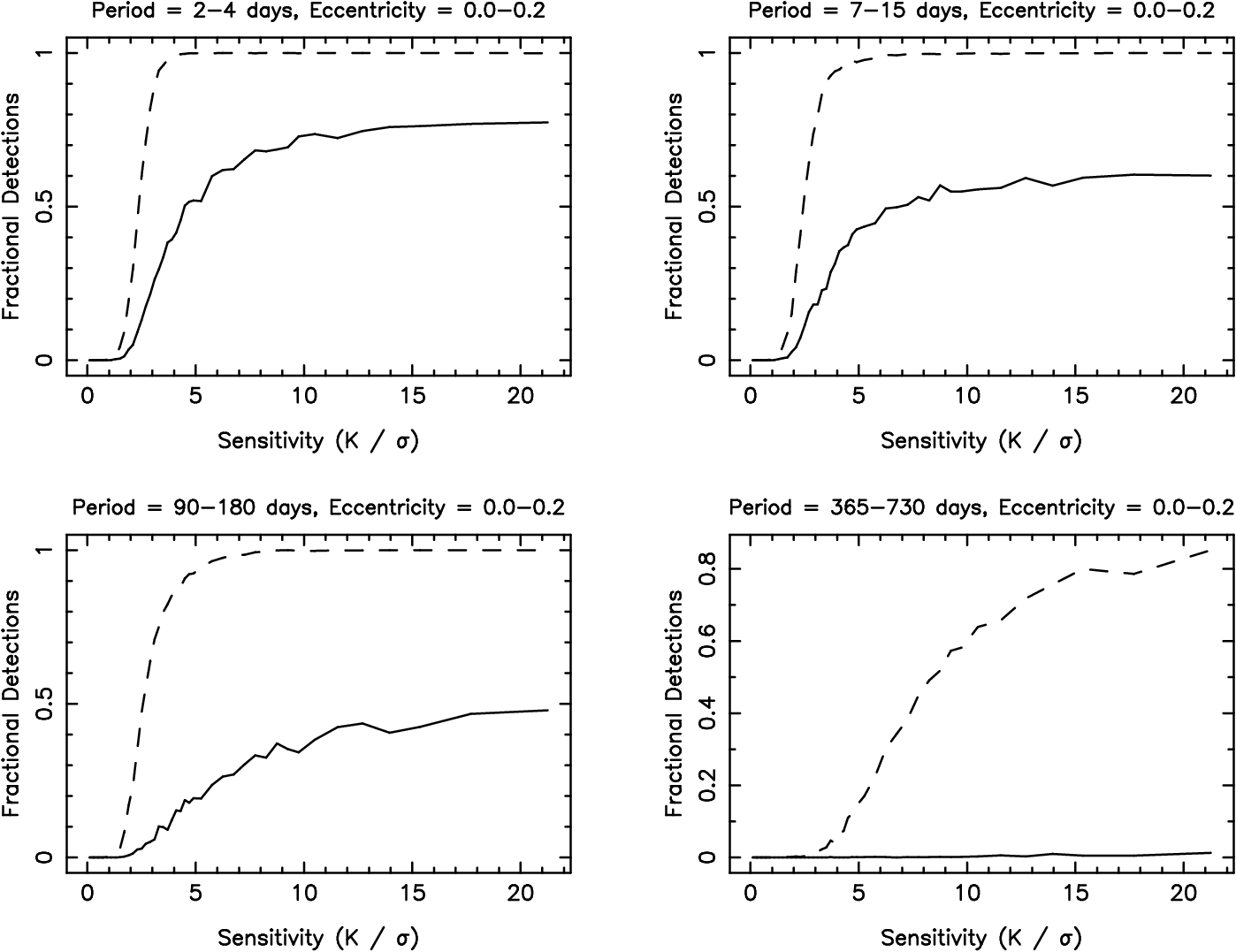} &
      \hspace{0.3cm}
      \includegraphics[angle=270,width=6.3cm]{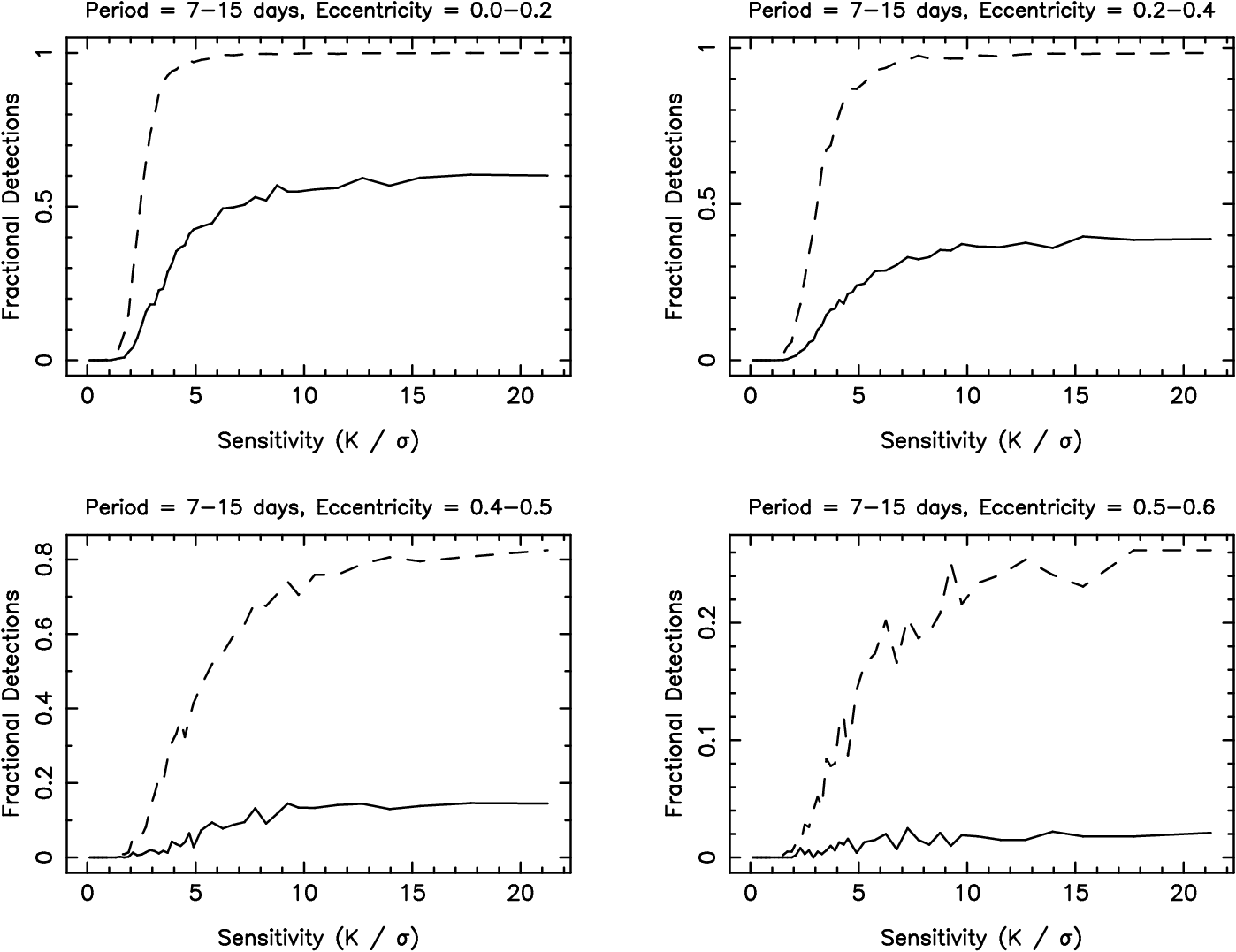} \\
    \end{tabular}
  \end{center}
  \caption{Detection efficiency results for the 888 cadence configuration, with
    the 4 plots on the left showing sensitivity variation with period, and the
    4 plots on the right showing sensitivity variation with eccentricity.}
\end{figure*}

A far superior plan is to use slightly more measurements spread over a
much longer time-scale. An example of this is an 18 month cadence with 33
measurements distributed in a cadence configuration described as
771111111000222222. The results of this simulation are shown in Figure 4 in
which it can be seen that the detection efficiency for both period and
eccentricity fare significantly better than in the 888 case.

\begin{figure*}
  \begin{center}
    \begin{tabular}{cc}
      \includegraphics[angle=270,width=6.3cm]{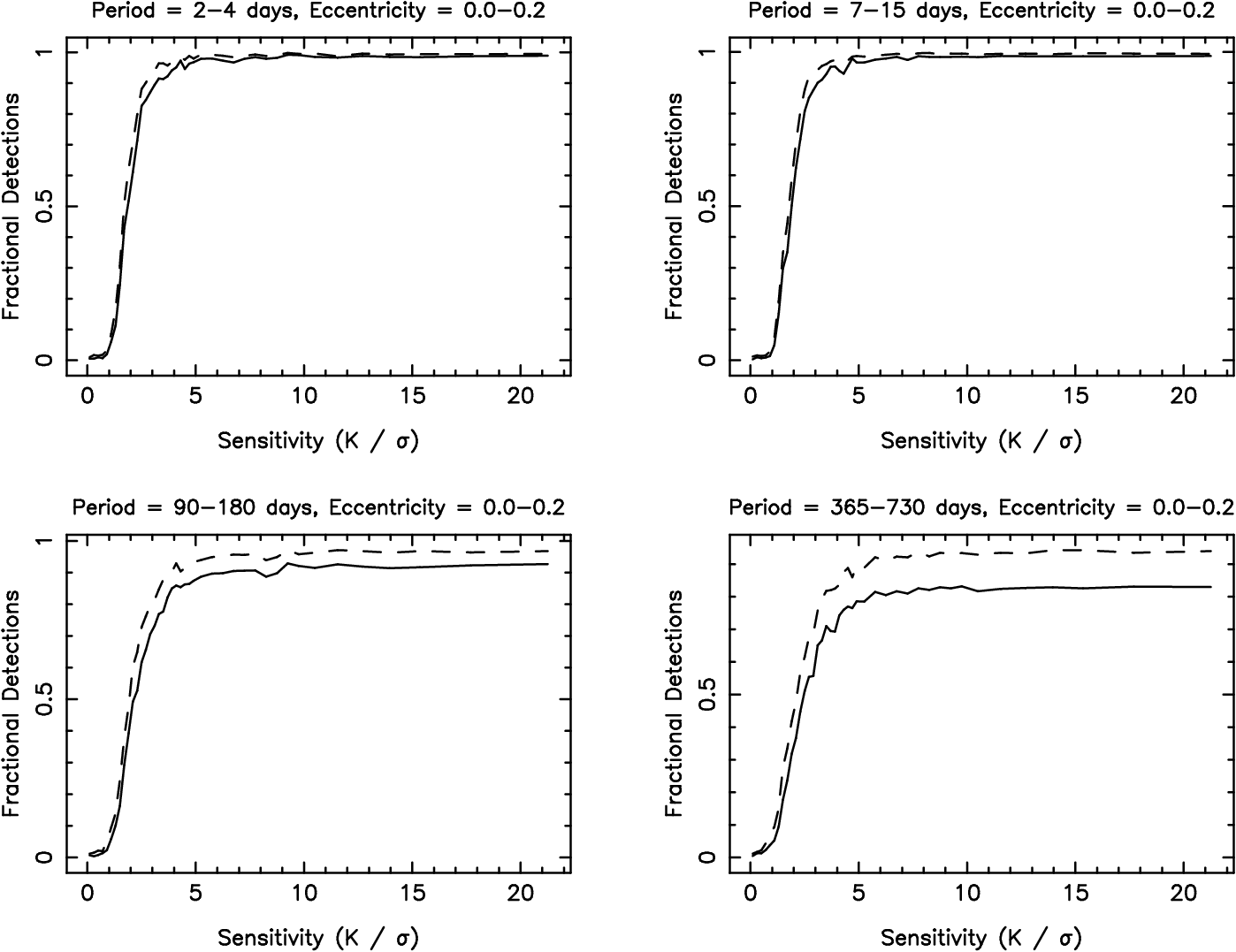} &
      \hspace{0.3cm}
      \includegraphics[angle=270,width=6.3cm]{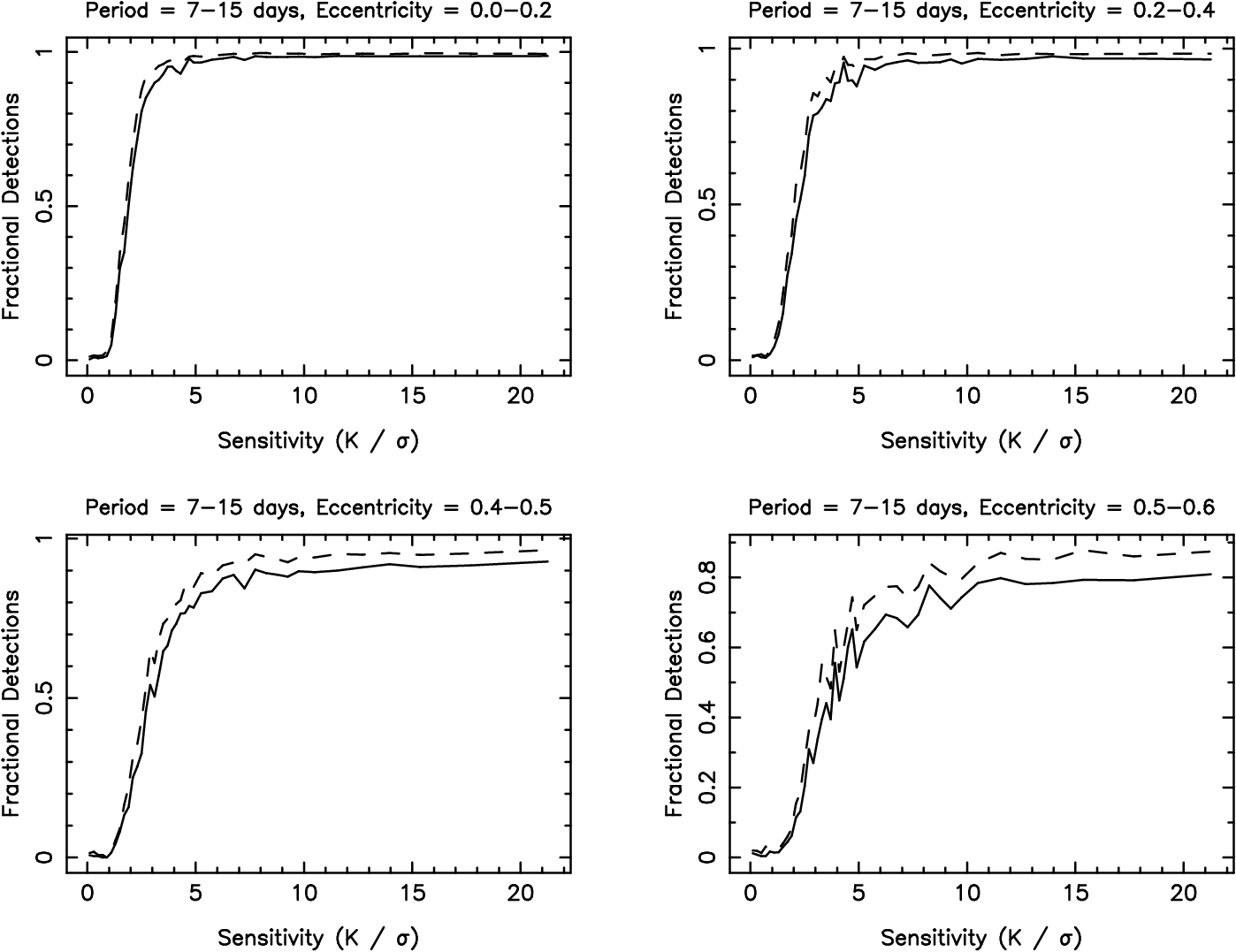} \\
    \end{tabular}
  \end{center}
  \caption{Detection efficiency results for the 771111111000222222 cadence
    configuration, with the 4 plots on the left showing sensitivity variation
    with period, and the 4 plots on the right showing sensitivity variation
    with eccentricity.}
\end{figure*}

A large number of cadence configurations have been investigated in this
manner. Given the parameter sensitivities derived from the cadence
simulations, we can now use the known distribution of exoplanetary
parameters to calculate the planet yield for each cadence. Table 1 shows
the planet yield predictions, unique and total detections, from a subset of
the cadences which, given uncertainties in stellar properties, provides a
useful comparison. Included in this table is a simulation in which the
measurements were distributed uniformly over a 36 month period, avoiding
monsoon seasons such as those experienced in Arizona.

\begin{table}
  \begin{center}
    \caption{Sample of cadence simulations performed with planet yields.}
    \begin{tabular}{cccc}\hline
{\bf measurements} & {\bf cadence}      & {\bf unique} & {\bf total} \\ \hline
15                 & 111111111000111111 &      31      &        41\\
24                 & 741111111000111111 &      88      &        94\\
33                 & 771111111000222222 &     173      &       200\\
33                 & 36 months uniform  &     184      &       187\\
    \end{tabular}
  \end{center}
\end{table}

There is a clear trade-off between the number of measurements and the
number of unique detections. The choice of cadence therefore largely
depends upon the amount of follow-up resources available. Figure 5 is a
cumulative histogram of the number of unique detections for a given cadence.
Beyond 30--35 data points, the fractional increase in unique detections
becomes negligible, therefore suggesting that it would be most useful for
increasing planet yield to change targets beyond this point.

\begin{figure}
  \hspace{2.5cm}
  \includegraphics[angle=270,width=8.2cm]{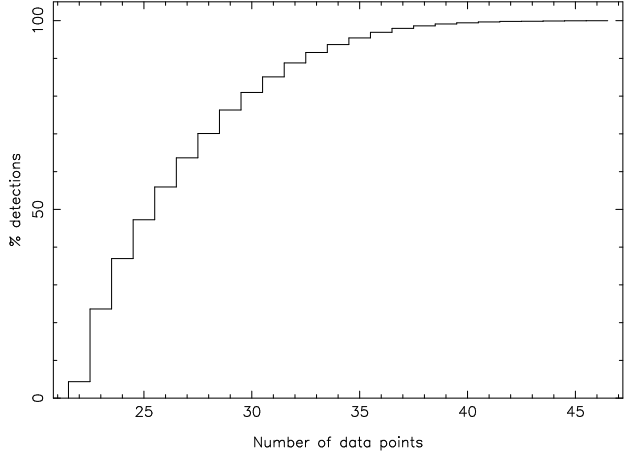}
  \caption{Cumulative histogram of the number of unique detections per
    measurement bin for a given cadence.}
\end{figure}

\section{Conclusions and Future Work}

These simulations show that the choice of observing cadence can have a
major impact on the exoplanetary parameter sensitivity. For example,
reducing the number of measurements from 33 to 15 has a devastating impact
on the planet yield. Furthermore, restricting the 33 measurements to 18
months rather than 36 months increases the sensitivity to short-period
planets. The detection of mid-high eccentricity planets are biased against
by the current algorithm, but this is being addressed by investigating the
inclusion of higher order fourier terms. Since there is a continuum of
possible cadence configurations, techniques to perform a more systematic
search of cadence ``parameter space'' are being developed to determine
optimal cadence solutions.

\end{document}